\documentclass[twocolumn,twoside,slac_two]{revtex4}
\usepackage{graphicx}
\usepackage{fancyhdr}
\newcommand{\gradi}{\ensuremath{^\circ}}
\newcommand{\dd}{\mathrm{d}}

\begin{document}

\title{Search for a neutrino emission from the Fermi Bubbles \\ with the ANTARES telescope}

%

\author{S. Biagi\footnote{simone.biagi@bo.infn.it}  on behalf of the ANTARES Collaboration}
\affiliation{Dipartimento di Fisica dell'Universit\`a  and INFN Sezione di Bologna, \\ Viale Berti Pichat 6/2, 40127 Bologna, Italy}

\begin{abstract}
ANTARES is the largest neutrino telescope in the Northern hemisphere. 
The main scientific goal is the search for cosmic neutrinos coming from galactic and extragalactic sources. Neutrinos are detected through the Cherenkov light emitted along the path of charged particles produced in neutrino interactions inside or in the vicinity of the detector. ANTARES is sensitive to all flavors though it is optimized for muon neutrinos. The detector has been taking data in its complete configuration since May 2008. 

Using data collected in the period 2007-2010, 
the first analysis devoted to the search for neutrinos from the Fermi Bubbles is presented. The Fermi Bubbles are characterized by gamma emission with a E$^{-2}$ spectrum and a relatively constant intensity all over the space. According to a proposed hadronic mechanism for this gamma-ray emission, the Fermi Bubbles can be a source of high-energy neutrinos. No evidence of a neutrino signal  is found in the ANTARES data. Therefore upper limits are calculated for neutrino fluxes with different energy cutoffs.
\end{abstract}

\maketitle

\thispagestyle{fancy}


\section{INTRODUCTION\label{sec:intro}}

\begin{figure}[t]
\includegraphics[width=53.6mm]{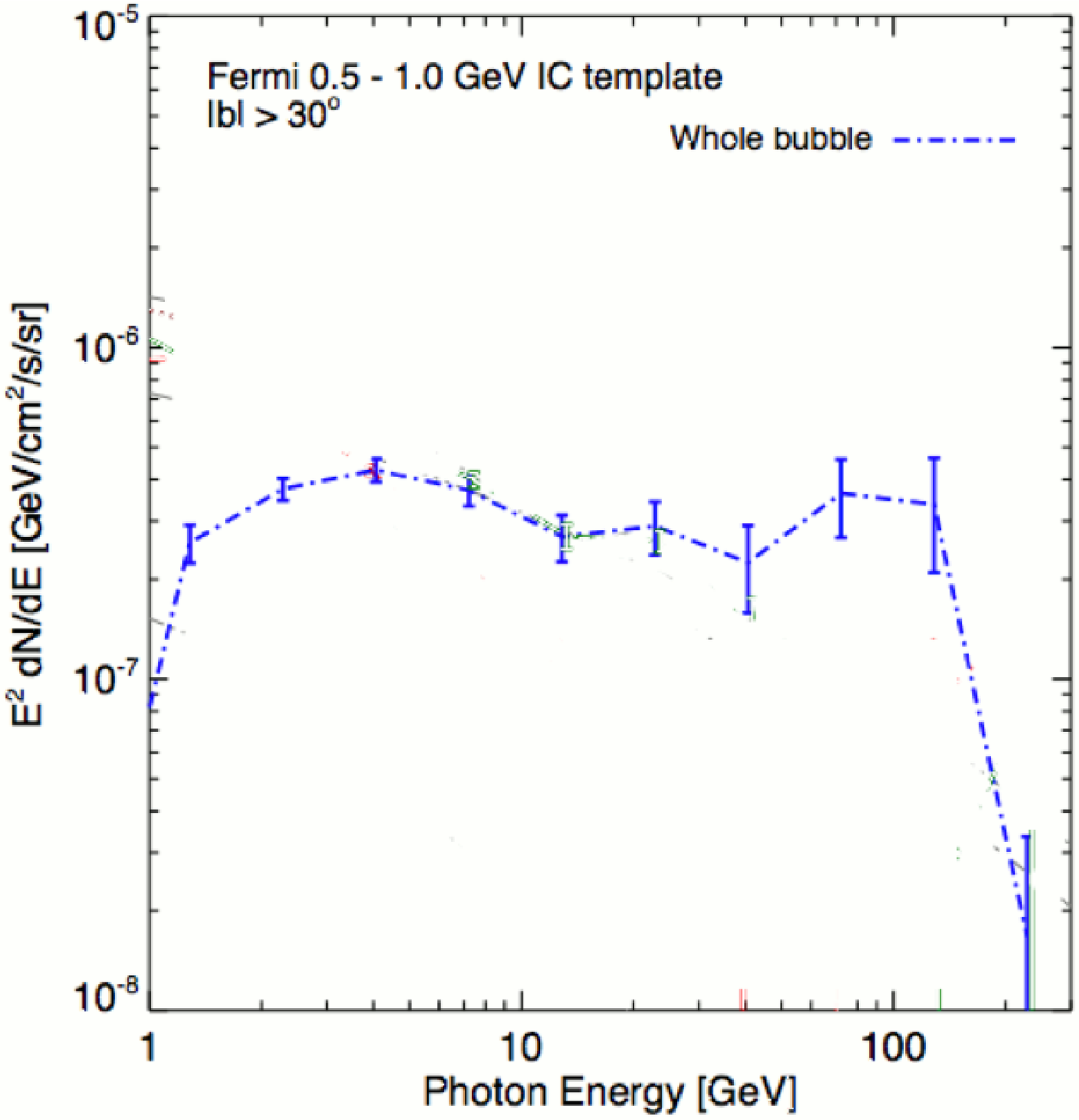}  \includegraphics[width=26.6mm]{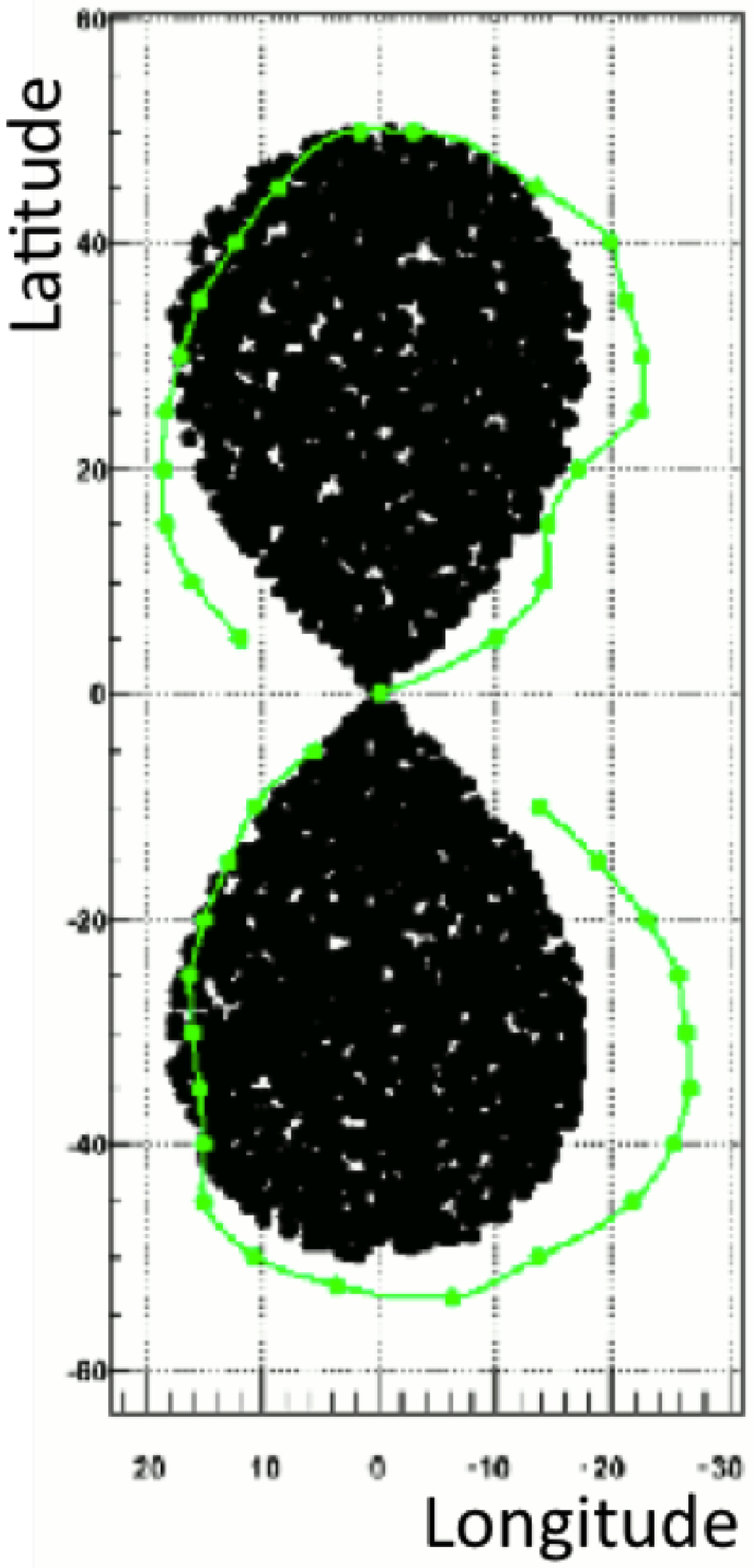}
\caption{Left: Gamma-ray flux of the FBs multiplied for $E^2$ taken from \cite{fb}. Right: Green line indicates  the shape of the FBs in galactic coordinates from \cite{fb}, black regions are  the approximation used in this analysis.} \label{fig:spectrum}
\end{figure}

The Fermi Bubbles (FBs) are extended regions characterized by gamma emission with a spectrum $\propto E^{-2}$  \cite{fb}. 
They cover $\sim 0.8$ sr in the sky and are centered around the Galactic Center, almost symmetrically with respect to the Galactic Plane; 
in  Fig. \ref{fig:spectrum} it can be seen that this emission has  a relatively constant intensity all over the range around the value  of $3\cdot10^{-7}$ GeV cm$^{-2}$ s$^{-1}$ sr$^{-1}$.

\begin{figure}[h]
\includegraphics[width=84mm]{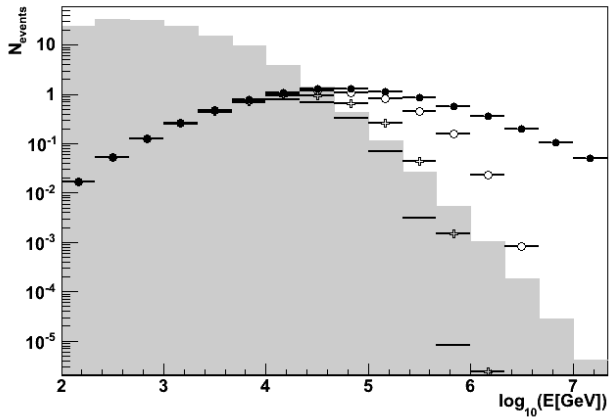}
\caption{Expected number of neutrino events from the FBs region corresponding to 4 years of data  as a function of the neutrino energy. The gray area represents the conventional  atmospheric neutrinos  and points are the neutrinos from the FBs: without energy cutoff (full circles), 500 TeV cutoff (empty circles), 100 TeV cutoff (crosses) and 50 TeV cutoff (black lines).} \label{fig:events_cutoffs}
\end{figure}

According to a proposed hadronic mechanism for gamma ray emission \cite{aharonian}, the FBs can be a source of high-energy neutrinos. From the measured gamma flux it is possible to derive the neutrino flux \cite{vissani}:
\begin{equation}\label{eq:ratio_flux}
\Phi_\nu \approx \frac{\Phi_\gamma}{2.5} 
\end{equation}
that results in
\begin{equation}\label{eq:flux}
E^2  \ \frac{\dd \Phi_\nu}{\dd E} \approx 1.2   \cdot10^{-7} \  \mathrm{GeV \  cm^{-2} \  s^{-1} \  sr^{-1}}
\end{equation}
According to \cite{aharonian}, the neutrino energy spectrum should present an exponential  cutoff  $\Phi \sim E^{-2} e^{-E/X}$. Four different values for the cutoff  are assumed in the following: no cutoff ($X=\infty$), 500 TeV, 100 TeV, and 50 TeV. 
In Fig. \ref{fig:events_cutoffs} the   number of expected neutrinos from the FBs for different energy cutoffs compared with the conventional atmospheric neutrino flux is shown as a function of the simulated neutrino energy. 

\section{THE ANTARES NEUTRINO TELESCOPE\label{sec:detector}}

\begin{figure}[t]
\includegraphics[width=80mm]{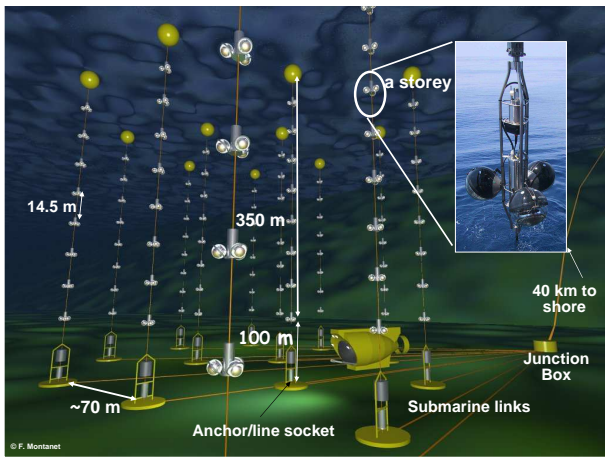}
\caption{The ANTARES neutrino telescope is a three-dimensional array of 885 photomultipliers distributed over 12 lines anchored on the seabed at distances of about 70 m from each other and tensioned by a buoy at the top of each line.} \label{fig:antares}
\end{figure}

The ANTARES neutrino telescope is located in the Mediterranean Sea close to the southern French coast of the city of Toulon  \cite{antares}. 885 photomultipliers tubes (PMTs) mounted on 12 strings are installed at a depth of 2500 meters (see Fig. \ref{fig:antares}) and detect the Cherenkov light emitted by ultra relativistic neutrino-induced muons along their path. The time and the charge collected by the PMTs (the \textit{hits}) are digitized and sent on-shore for triggering and  storing on disk.  The collected hits are used to reconstruct the direction of the primary neutrino and to estimate its energy. The track reconstruction algorithm  is based on a likelihood fit that uses a detailed parametrization of the probability density function for the photon arrival times and gives as  outputs  the  position and direction of the muon track, the information on the number of hits ($N_{hit}$) used for the reconstruction, and  a quality parameter $\Lambda$.  The neutrino-induced tracks are selected as ``upgoing'' to reject the dominant background of atmospheric muons;   cutting on the   $\Lambda$ quality  parameter is possible to reduce the contamination of mis-reconstructed as upgoing atmospheric muons  to the level of few percent.

\section{ON/OFF ZONES APPROACH\label{sec:on_off}}

As it can be seen in Fig. \ref{fig:events_cutoffs}, the expected event flux  from the FBs is  several order of magnitude lower than the atmospheric neutrino flux, 
that represents an irreducible background.
A discrimination will be done on the basis of the  energy 
and a correct estimation of the atmospheric background is fundamental in this analysis. 

MonteCarlo (MC) $\nu_\mu$ events generated to reproduce the atmospheric neutrinos are weighted using the so-called conventional ``Bartol'' flux \cite{bartol}. Atmospheric neutrinos are produced by the interaction of high-energy cosmic rays in the atmosphere. The uncertainties on the flux of  atmospheric neutrinos are at the level of   25$\div$30\% due to the lack of   measurements of the cosmic ray fluxes at high energies  and to the uncertainties on the cross sections of cosmic rays with the light atoms in the upper atmosphere.

In addition systematic uncertainties about the detector simulation (absorption length of light in sea water, PMT efficiency) can make the data/MC comparison inefficient for revealing the possible signal. The solution adopted is to estimate the background in the FBs zone directly from data, defining different zones in the sky with the same coverage and visibility.
 The main idea of this method is to compare the  measurement done in the FBs area (ON zone) with a background estimated from  data itself looking into an  area   outside the FBs (OFF zone). 
 
\begin{figure}[t]
\includegraphics[width=82mm]{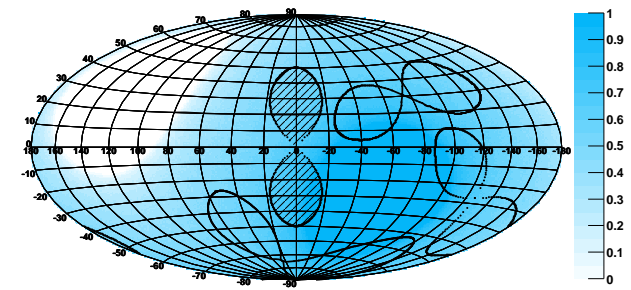}
\caption{Position in galactic coordinates of the FBs region (the ON zone) and of the three chosen regions with the same visibility (the OFF zones) superimposed on the total ANTARES visibility.} \label{fig:visibility}
\end{figure}

 In local coordinates, the FBs move in the sky. A zone with the same position but shifted in time follows the FBs zone. A proper choice of the time shift avoids an overlapping of the zones, allowing an unambiguous definition of the OFF zones. 
 The expected  number of background events  is proportional to the efficiency of the detector, that is a function of the local coordinates only.
 In this analysis, three OFF zones (Fig. \ref{fig:visibility}) are identified corresponding to 6, 12 and 18 hours time shifts from the FBs (ON zone); moreover this choice  reduces the background uncertainty.

\begin{figure*}[t]
\centering
\includegraphics[width=83mm]{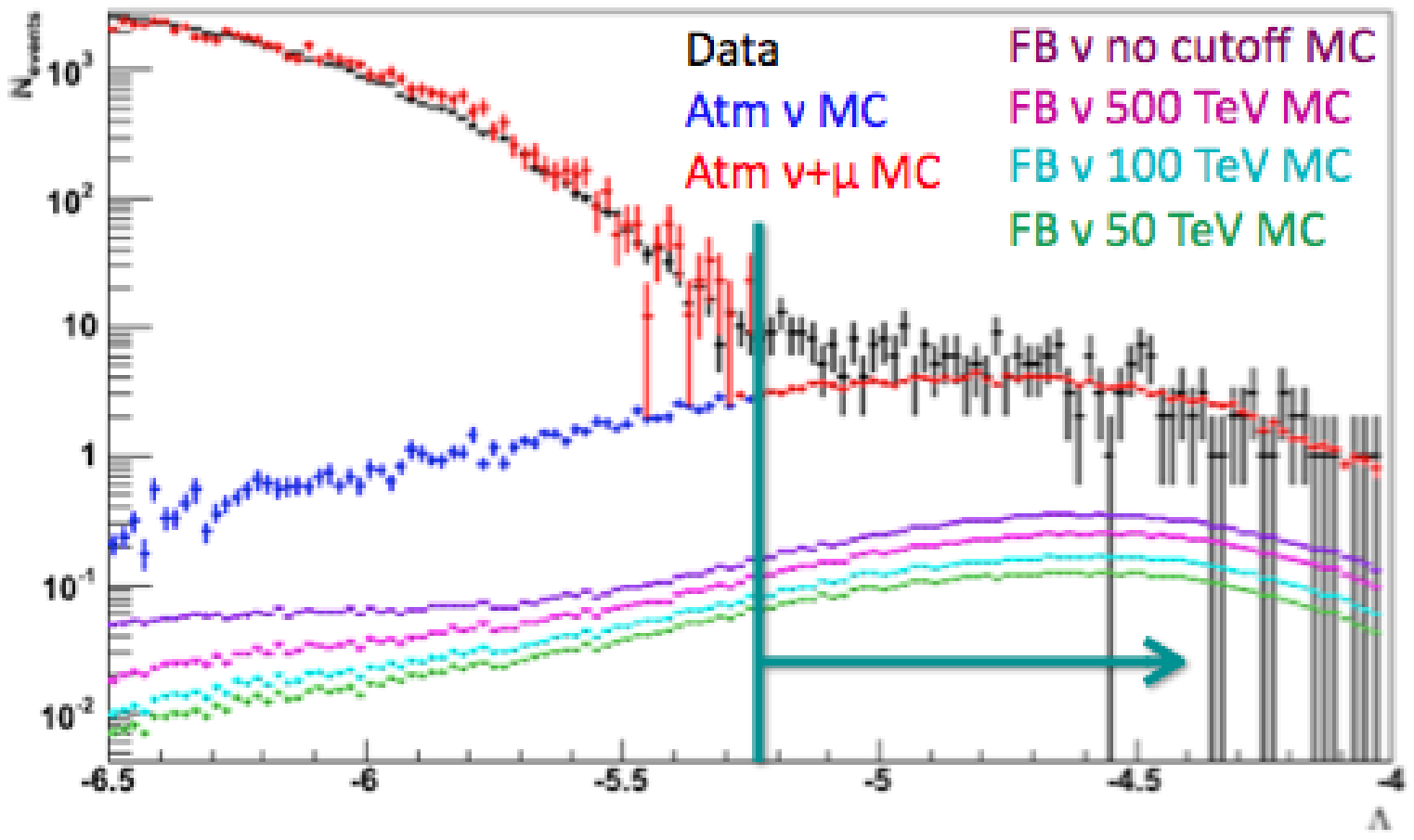}  \includegraphics[width=86mm]{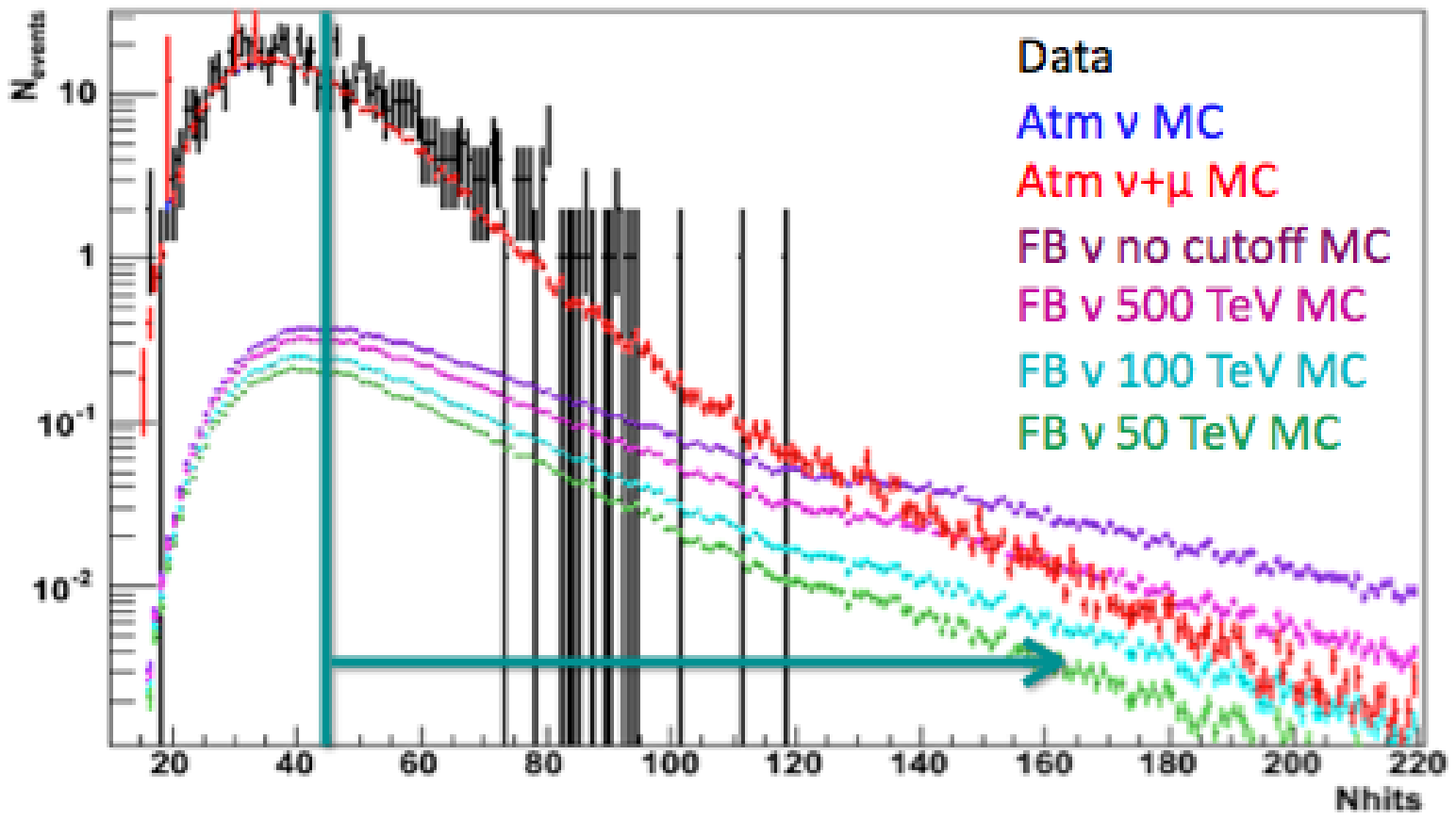}
\caption{Left: Reconstruction quality parameter $\Lambda$ distributions for data and MC. Events with the best $N_{hit}$ are selected. Several models for the neutrino spectrum are displayed.  Right: Data and MC events as a function of the total number of hits.   The cut that defines the high energy region is obtained with the MRF procedure -- see text. } \label{fig:lambda}
\end{figure*}

\section{SENSITIVITY OPTIMIZATION AND BLINDING PROCEDURE\label{sec:sensitivity}}

Background in this analysis is estimated directly from data. 
To verify that the OFF zones have the same visibility, the data event rate is compared with MC expectation;  
a conservative systematic error of 3\% is evaluated to account for different visibilities of the zones.
The optimization of the analysis cuts is done using  MC  simulations in order to maximize the sensitivity to a FBs neutrino signal and to reject atmospheric muons and atmospheric neutrinos (blinded analysis). 

Cuts are optimized according to the prescriptions of the Model Rejection Factor (MRF) procedure \cite{mrf}. 
This method uses the Feldman and Cousins statistics    to calculate  upper limits  at 90\% c.l. \cite{feldman}, for an ensemble of hypothetical experiments with no signal and a background with a Poissonian probability of occurrence.  The MRF represents the average upper limit that can be obtained in case of no discovery, i.e. the sensitivity of ANTARES to the assumed signal flux. The minimization of the MRF produces the best cut on the chosen parameters, in our case the reconstruction quality parameter $\Lambda$ and the number of hits used in reconstruction, $N_{hit}$.

The rejection of downward-going atmospheric muons is achieved cutting on the $\Lambda$  parameter; the energy estimator,  $N_{hit}$, can provide a discrimination between atmospheric and signal neutrinos (Fig. \ref{fig:lambda}).
Neutrino tracks reconstructed with only one line and with an angular error greater than $1\gradi$ are rejected. 

For simplicity, a unique event selection is chosen for the four tested spectrum models ($\Lambda>-5.24$ and $N_{hit}>44$), obtained with the MRF minimization of the 100 TeV cutoff signal.
The corresponding sensitivity for the considered energy cutoffs is reported in Table \ref{tab:mrf}.


\begin{table}[t]
\begin{center}
\caption{MRF and sensitivity to a neutrino flux $\propto E^{-2}  e^{-E/X}$ (in units of 10$^{-7}$ GeV cm$^{-2}$ s$^{-1}$ sr$^{-1}$) for different energy cutoffs. $\Lambda$ and $N_{hit}$ cuts are optimized to get the best sensitivity.}
\begin{tabular}{|c|c|c|}
\hline \textbf{ Cutoff [TeV] } & \textbf{ \ \ MRF \ \ } & \textbf{ Sensitivity }  \\
\hline no cutoff & 2.75 & 3.30 \\
\hline 500 & 3.79 & 4.55 \\
\hline 100 & 5.74 & 6.89 \\
\hline 50 & 7.58 & 9.09 \\
\hline
\end{tabular}
\label{tab:mrf}
\end{center}
\end{table}

\section{RESULTS AND UPPER LIMIT\label{sec:results}}

Data are unblinded in the ON region searching for an excess of events in comparison to the measured background in the three OFF regions. 75 events are observed in the ON region with $90\pm5$(stat)$\pm3$(sys) background events evaluated through an average of the OFF regions. 
No evidence of a neutrino signal from the FBs region is found.

\begin{figure*}[t]
\centering
\includegraphics[width=110mm]{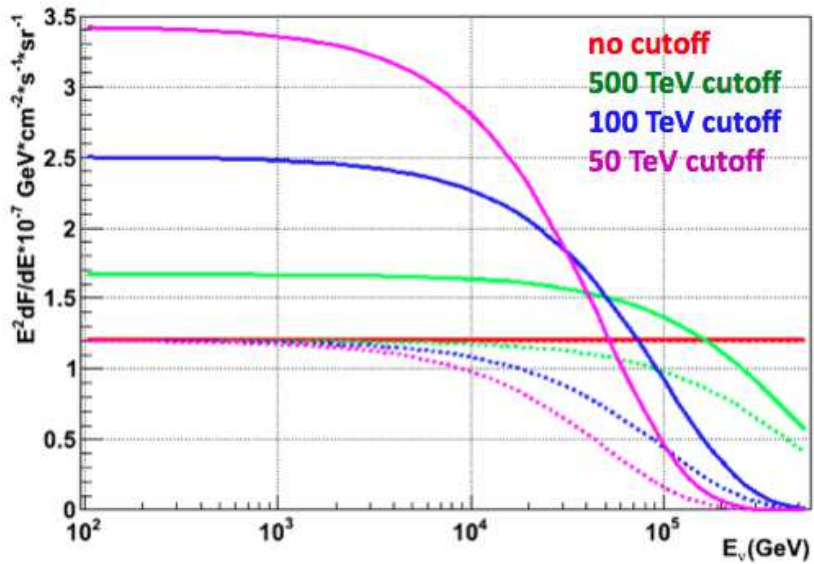}
\caption{ANTARES upper limits at a  90\% c.l.  for a $E^{-2}$ neutrino flux from the Fermi Bubbles with different energy cutoffs. Theoretical predictions normalized at the value of eq. \ref{eq:flux} are plotted with dotted lines. \\ \tiny{   \ \ }} \label{fig:limits}
\end{figure*}

Upper limits at 90\% c.l. are computed using the Feldman and Cousins recipe with 75 observed events and 90 background events. In addition, a systematic error of $^{+15\%}_{-6\%}$ for data/MC  comparison is taken into account in the limit calculation. Due to a negative fluctuation of background in the ON zone, the quoted upper limits are lower than the ANTARES sensitivity to neutrino fluxes. 
A comparison of these limits with the assumed theoretical models is presented  in Fig. \ref{fig:limits}; the limits for the most optimistic cutoffs are very close to the expected fluxes.

Very soon, data collected in 2011 will be added  to increase the statistical significance of the analysis. 
Furthermore, ANTARES is currently  developing a method to combine various observables through a likelihood approach in order to increase the energy resolution. 
An Artificial Neural Network (ANN)  is  used for the mapping of the likelihood between the chosen observables and the energy: about 50 parameters are used as input of the ANN (hits from various triggers, parameters from tracking algorithms, etc)  \cite{jutta}.  After the selection of the input parameters, the ANN  must be trained with a sample of    neutrino events generated with    MC simulations; the final achieved energy resolution is about 0.3 of the  logarithm of the reconstructed energy  in the region between 1 TeV and  300 TeV.
The ANN will be used in the final version of this analysis, improving the sensitivity of ANTARES.

\end{document}